\documentclass[prd,preprint,superscriptaddress,amsmath,amssymb,nofootinbib]{revtex4}
\pdfoutput=1

\usepackage{graphicx,color,slashed}

\def\be{\begin{eqnarray}}
\def\ed{\end{eqnarray}}

\begin{document}


\title{\bf \Large The decay $A^0\to h^0 Z^{(*)}$ in the inverted hierarchy scenario and its detection prospects at the Large Hadron Collider}

\author{A.G. Akeroyd}
\email{a.g.akeroyd@soton.ac.uk}
\affiliation{School of Physics and Astronomy, University of Southampton,
Highfield, Southampton SO17 1BJ, United Kingdom}

\author{S. Alanazi}
\email{swa1a19@soton.ac.uk; SWAlanazi@imamu.edu.sa}
\affiliation{School of Physics and Astronomy, University of Southampton,
Highfield, Southampton SO17 1BJ, United Kingdom}
\affiliation{Physics Department, Imam Mohammad Ibn Saud Islamic University (IMISU), P.O. Box 90950, Riyadh, 11623, Saudi Arabia}

\author{Stefano Moretti}
\email{S.Moretti@soton.ac.uk; stefano.moretti@physics.uu.se}
\affiliation{School of Physics and Astronomy, University of Southampton,
Highfield, Southampton SO17 1BJ, United Kingdom}
\affiliation{Department of Physics and Astronomy, Uppsala University, Box 516, SE-751 20 Uppsala, Sweden}

\date{\today}

\begin{abstract}
\noindent
Searches are being carried out at the Large Hadron Collider (LHC) for the decay of the CP-odd scalar ($A^0$) in Two-Higgs-Doublet Models (2HDMs) with Natural Flavour Conservation (NFC) in the channel
$A^0\to h^0 Z$ (with $m_{h^0}=125$ GeV and $Z$ on-shell).
In the absence of any signal, limits on the parameter space of $[\tan\beta, \cos(\beta-\alpha), m_{A^0}]$ in each 2HDM are derived for $m_{A^0} > 225$ GeV. 
In this work we consider the scenario of inverted
hierarchy with $m_{h^0}<125$ GeV and $m_{H^0}=125$ GeV in which the decay $A^0\to h^0 Z^{(*)}$ (i.e. including the case of an off-shell $Z$) can have a large branching ratio in the 2HDM (Type I) for $m_{A^0}<225$ GeV. We
calculate the signal cross section $\sigma(gg\to A^0)\times {\rm BR}(A^0\to h^0Z^{(*)})\times {\rm BR}(h^0\to b\overline b)$ in the 2HDM (Type I) with NFC and compare its magnitude with the cross section for the case of normal hierarchy ($m_{h^0}=125$ GeV) that is currently being searched for at the LHC. For the experimentally unexplored region $m_{A^0} < 225$ GeV
it is shown that the above cross section for signal events in the scenario of inverted hierarchy can be of the order of a few picobarns.
Such sizeable cross sections are several orders of magnitude larger than the cross sections for the case of normal hierarchy, thus motivating an extension of the ongoing searches for $A^0\to h^0 Z^{(*)}$
to probe the scenario of inverted hierarchy.

\end{abstract}

\maketitle

\section{Introduction}
\noindent
The discovery in the year 2012 of a new particle with a mass of around 125 GeV by the ATLAS and CMS collaborations of the Large Hadron Collider (LHC)
\cite{Aad:2012tfa,Chatrchyan:2012xdj} has led to increasingly precise measurements of its properties in the last ten years.
To date, all measurements of the 125 GeV state are in very good agreement 
(within experimental error) with the predicted properties of the Higgs boson of the 
Standard Model (SM) with a mass of 125 GeV. 
Five decay channels ($\gamma\gamma$, $ZZ$, $W^+W^-$, $\tau^+\tau^-$, and $b\overline b$) 
have now been observed with a statistical significance of greater than 
$5\sigma$ (e.g. see \cite{ATLAS:2018kot}). Evidence for the decays to $\mu^+\mu^-$ and $Z\gamma$ is currently at the $2\sigma$ level, and observation of
these channels
with a statistical significance of $5\sigma$ is likely by the end of the operation of the High Luminosity LHC (HL-LHC).
In addition, each of the four main
production mechanisms (gluon-gluon fusion, vector boson $(W/Z)$ 
fusion, associated production with a vector boson, and 
associated production with top quarks) have been measured for at least one of the above decay channels,
with no significant deviation from the predicted cross-sections of the SM Higgs boson.
Measurements of all the above cross sections and branching ratios (BRs) with the full Run II data (139 fb$^{-1}$ at $\sqrt s=13$ TeV) have been combined
to show a signal strength (i.e. cross section times BR, averaged over all channels) relative to that of the SM Higgs boson of $1.02^{+0.07}_{-0.06}$ \cite{CMS:RunII} (CMS) and $1.06\pm 0.06$ \cite{ATLAS:RunII} (ATLAS).

Whether or not the observed 125 GeV boson is the 
(solitary) Higgs boson of the SM is still an issue to be clarified experimentally.
It is possible that the 125 GeV boson is the first scalar to be discovered from an extension of the SM that contains a non-minimal Higgs 
sector e.g. the scalar potential contains additional scalar isospin doublets 
and/or other representations such as scalar isospin singlets/triplets. A much studied example is the non-supersymmetric Two Higgs Doublet Model (2HDM)
 \cite{Lee:1973iz,Gunion:1989we,Branco:2011iw,Wang:2022yhm}, in which the scalar potential of the SM
contains two $SU(2)_L\otimes U(1)_Y$ isospin doublets instead of just one.
The SM has various shortcoming such as i) an absence of neutrino mass, 
ii) an absence of a dark matter candidate, and iii) insufficient CP violation for baryogenesis. These issues (and others) are often solved in extensions of the SM
that contain additional scalars. Many models with a non-minimal Higgs sector predict a SM-like scalar in part of the model's parameter space. In the aforementioned
2HDM there is an "alignment limit" in which one of the CP-even scalars has properties that exactly match those of the Higgs boson of the SM. This alignment is naturally obtained
if only one of the CP-even scalars remains light (of the order of the electroweak scale) while all other scalars have masses that are much larger. The alignment can also be realised if
all scalars are of the order of the electroweak scale ("alignment without decoupling") and it is on this scenario that we will focus.

If the 125 GeV boson is the first scalar to be discovered from a non-minimal Higgs sector then
future measurements (e.g. with larger integrated luminosity at the LHC and/or at a future $e^+e^-$ collider) of its various production cross sections and BRs
might start to show deviations from the values for the SM Higgs boson. Moreover, 
enlarged Higgs sectors contain additional neutral scalars and/or charged scalars ($H^\pm$), and such particles are being actively searched for at the LHC.
In 2HDMs there are two CP-even scalars $h^0$ and $H^0$ (with $m_{h^0} < m_{H^0}$), a pair of charged scalars $H^+$ and $H^-$ and
a neutral pseudoscalar Higgs boson $A^0$, which is CP-odd. 

The discovered 125 GeV boson has been shown to be CP-even and in the context of a 2HDM it would be interpreted as being either $h^0$ (called "normal hierarchy", NH) or $H^0$ (called "inverted hierarchy", IH).
The CP-odd $A^0$ does not have tree-level couplings to the
 gauge bosons of the weak interaction $(W^\pm, Z$) and has a different phenomenology to both $h^0$ an $H^0$.
We shall focus on the prospects of discovering an $A^0$ from a 2HDM at the LHC via its decay $A^0\to h^0Z^{(*)}$. In the context of NH
one has $m_{h^0}=125$ GeV and the current searches at the LHC for $A^0\to h^0Z$ (assuming an on-shell $Z$) are only carried out for this NH scenario and for the specific case of $m_{A^0}>$225 GeV. In this work we consider the case of IH in which
$m_{h^0}$ can be significantly lighter than 125 GeV. It will be shown that the number of signal events for $A^0\to h^0Z^{(*)}$ can be considerably larger than
in NH for the experimentally unexplored region of $m_{A^0}<$ 225 GeV, and the current experimental searches would need to be modified in order to probe this scenario.

This work is organised as follows. In section II the various 2HDMs are introduced.
In section III the phenomenology of $A^0$ at the LHC is presented, and in section IV the current searches for $A^0\to h^0 Z$ at the LHC are summarised.
Our numerical results for the cross section for $A^0\to h^0 Z^{(*)}$ events in the IH scenario are given in section V, and conclusions are contained in
section VI.

\section{The Two Higgs Doublet Model (2HDM)}
\noindent
The SM has one complex scalar isospin doublet $(I=1/2)$ with hypercharge $Y=1$, in which the real part of the neutral scalar field obtains a vacuum expectation value ($v$).
The presence of $v$ leads to the spontaneous breaking of the $SU(2)_L\otimes U(1)_Y$ local gauge symmetry to a $U(1)_Q$ local gauge symmetry, and provides mass to the $W^\pm, Z$ (via the kinetic energy term of the scalar fields)
and charged fermions (via the Yukawa couplings). Such a mechanism for the generation of mass is called the "Higgs mechanism", and a CP-even physical scalar particle (a "Higgs boson", $h^0$) is predicted. In the context of the SM this Higgs boson $h^0$ has now been found with a mass of around 125 GeV.
 The Higgs mechanism can also be implemented using two complex scalar doublets in which there are now two vacuum expectation values ($v_1$ and $v_2$), and such a model is called the 2HDM \cite{Lee:1973iz,Gunion:1989we,Branco:2011iw,Wang:2022yhm}.
Supersymmetric (SUSY) versions of the SM require two complex scalar doublets \cite{Djouadi:2005gj}, but the 2HDM has also been well-studied as a minimal (and non-SUSY) extension of the SM.
  After "electroweak symmetry breaking" (EWSB) there are five physical Higgs bosons instead of the one CP-even Higgs boson $h^0$ of a one-scalar doublet model. 
 In the context of a 2HDM the 125 GeV boson that was discovered at the LHC is interpreted as being either
  $h^0$ (NH) or $H^0$ (IH), with couplings very close to those of the SM Higgs boson.

Enlarging the scalar sector of the SM can conflict with experimental data. A strong suppression of "Flavour Changing Neutral Currents" (FCNCs) that are predicted in any 2HDM is a stringent constraint
on its structure.
In general, the Yukawa couplings in a 2HDM are not flavour diagonal. Such FCNCs lead to interactions that change quark flavour (such as a vertex $h^0b\overline s$), 
which must be highly suppressed in order to respect experimental limits on the phenomenology of quarks.
A particularly elegant suppression mechanism of FCNCs in 2HDMs (the "Paschos-Glashow-Weinberg theorem" or "Natural Flavour Conservation" (NFC) \cite{Glashow:1976nt}) is to require that the Lagrangian respects certain discrete symmetries ($Z_2$ symmetries).
Such symmetries enforce that a given flavour of charged fermion receives its mass from just one vacuum expectation value, leading
to the elimination of FCNC processes at the tree-level.

The  most general scalar potential of a 2HDM that is invariant under the $SU(2)_L\otimes U(1)_Y$ local gauge symmetry and which only softly breaks (via the $m^2_{12}$ terms) an appropriate $Z_2$ symmetry (imposed to avoid FCNCs) is as 
follows \cite{Gunion:1989we,Branco:2011iw}:
 \begin{eqnarray}
        V(\Phi _{1}\Phi _{2})  =  m_{11}^{2}\Phi _{1}^{\dagger }\Phi _{1}+m_{22}^{2}\Phi _{2}^{\dagger }\Phi _{2}-  m_{12}^{2}(\Phi _{1}^{\dagger }\Phi _{2}+\Phi _{2}^{\dagger }\Phi _{1})+  \frac{ \lambda _{1}}{2}(\Phi _{1}^{\dagger }\Phi _{1})^{2}+\\ \nonumber
         \frac{ \lambda _{2}}{2}(\Phi _{2}^{\dagger }\Phi _{2})^{2}+  \lambda _{3}\Phi _{1}^{\dagger }\Phi _{1}\Phi _{2}^{\dagger }\Phi _{2}+
          \lambda _{4}\Phi _{1}^{\dagger }\Phi _{2}\Phi _{2}^{\dagger }\Phi _{1}+\frac{ \lambda _{5}}{2}[(\Phi _{1}^{\dagger }\Phi _{2})^{2}+(\Phi _{2}^{\dagger }\Phi _{1})^{2}]\,,
 \end{eqnarray}
with $\Phi _{i}=\binom{\Phi _{i}^{\dotplus }}{\frac{(\upsilon _{i}+\rho _{i}+i\eta _{i})}{\sqrt{2}}},  \:{\rm and} \; i=1,2$. \\
In general, some of the parameters in the scalar potential can be complex and thus they can be sources of CP violation. We consider a simplified scenario by taking all parameters to be real, as is often done
in phenomenological studies of the 2HDM.
The scalar potential then has 8 real independent parameters: $m^2_{11}$, $m^2_{22}$, $m^2_{12}$, $\lambda _{1}$, $\lambda _{2}$, $\lambda _{3}$, $\lambda _{4}$, and $\lambda _{5}$.
These parameters determine the masses of the Higgs bosons and their couplings to fermions and gauge bosons. However, it is convenient to work with different independent parameters which are more directly related to physical observables.
A common choice is: $m_{h^0}$, $m_{H^0}$, $m_{H^\pm}$, $m_{A^0}$, $\upsilon_1$, $\upsilon_2$, $m^2_{12}$ and $\sin(\beta-\alpha)$. The first four parameters are the masses of the physical Higgs bosons.
The vacuum expectation values $\upsilon_1$ and $\upsilon_2$ are the values of the neutral CP-even fields in $\Phi_1$ and $\Phi_2$ respectively at the minimum of the scalar potential:
\begin{equation}
    \left<\Phi _{1} \right> =\frac{1}{\sqrt{2}}\binom{0}{\upsilon _{1}}\, ,\, \, \, \, \left<\Phi _{2} \right> =\frac{1}{\sqrt{2}}\binom{0}{\upsilon _{2}}\,.
\end{equation}
The parameter $\beta$ is defined via $\tan\beta=\upsilon_2/\upsilon_1$, and the angle $\alpha$ determines the composition of the CP-even mass eigenstates $h^0$ and $H^0$ in terms of the original 
neutral CP-even fields that are present in the isospin doublets $\Phi_1$ and $\Phi_2$. Of these 8 parameters in the scalar potential, 2 have now been measured. After EWSB in a 2HDM, the mass of the $W^\pm$ boson is given by
$m_W=gv/2$, with $ \upsilon=\sqrt{\upsilon^{2}_{1}+\upsilon^{2}_{2}}\simeq 246$ GeV. Hence only one of $\upsilon_1$ and $\upsilon_2$ is independent, and so $\tan\beta=\upsilon_2/\upsilon_1$ is taken as
an independent parameter. As mentioned earlier, in a 2HDM the discovered 125 GeV boson is taken to be $h^0$ or $H^0$ and thus either $m_{h^0}=125$ GeV (NH) or $m_{H^0}=125$ GeV (IH).
The remaining 6 independent parameters in the 2HDM scalar potential are: $m_{H^\pm}$, $m_{A^0}$, $m^2_{12}$, $\tan\beta$, $\sin(\beta-\alpha)$ and one of $[m_{h^0}, m_{H^0}]$. In the NH scenario $m_{H^0}>125$ GeV
and in the IH scenario $m_{h^0}<125$ GeV. In this work we shall be focussing on the IH scenario and the phenomenology of $A^0$.

As mentioned above, the  masses of the pseudoscalar $A^0$ and the charged scalars $H^{\pm}$ are independent parameters, and in terms of the original parameters in the
scalar potential are given by:
   \begin{equation}
       \begin{aligned}
         m_{A^0}^{2} & = \left [\frac{ m_{12}^{2}}{\upsilon _{1}\upsilon _{2}}  -2\lambda_{5}\right ](\upsilon _{1}^{2}+\upsilon _{2}^{2})\,,\\ m_{H^{\pm }}^{2} & = \left [\frac{ m_{12}^{2}}{\upsilon _{1}\upsilon _{2}} -\lambda_{4} -\lambda_{5}\right ](\upsilon _{1}^{2}+\upsilon _{2}^{2}) = \left [ m_{A}^{2}  +\upsilon (\lambda_{5}-\lambda_{4})\right ]  \,.
       \end{aligned}
   \end{equation}
  From these equations it can be seen that the mass difference between $m_{A^0}$ and $m_{H^\pm}$ depends on $\lambda_5-\lambda_4$. In our numerical analysis we
  shall be taking $m_{A^0}=m_{H^\pm}$ in order to satisfy more easily the constraints from electroweak precision observables ("oblique parameters"), and this corresponds to $\lambda_5=\lambda_4$.
  For the masses of the CP-even scalars we take $m_{H^0}=125$ GeV, and $m_{h^0} <125$ GeV (IH scenario). 
     
  There are four distinct types of 2HDM with NFC which differ in how the two doublets are coupled to the charged fermions. These are called: Type I, Type II, Lepton Specific and Flipped \cite{Barger}.
 The phenomenology of all four models has been studied in great detail.
 The Lagrangian in a 2HDM that describes the interactions of $A^0$ with 
the fermions (the Yukawa couplings) can be written as follows \cite{Branco:2011iw}:
\begin{equation}
{\cal L}^{yuk}_{A^0} =\frac{i}{v}\left(y^d_{A^0} m_d A^0 \overline d \gamma_5 d + y^u_{A^0} m_u A^0 \overline u \gamma_5 u+  y ^\ell_{A^0} m_\ell  A^0 \overline\ell \gamma_5 \ell \right)\,.
\label{yukawa}
\end{equation}
In eq.~(\ref{yukawa}) it is understood that $d$ refers to the down-type quarks ($d$, $s$, $b$), $u$ refers to the up-type quarks ($u$, $c$, $t$) and $\ell$ refers to the
charged leptons ($e$, $\mu$, $\tau$) i.e. there are three terms of the form $y^d_{A^0} m_d \overline d \gamma_5 d$.
 In Table \ref{2HDMAcoup} the couplings $y^d_{A^0}$, $y^u_{A^0}$, and $y^\ell_{A^0}$  of $A^0$ to the charged fermions in each of these four models
are displayed.
\begin{table}[h]
\begin{center}
\begin{tabular}{|c||c|c|c|}
\hline
& $y^d_{A^0}$ &  $y^u_{A^0}$ &  $y^\ell_{A^0}$ \\ \hline
Type I
&  $-\cot\beta$ & $\cot\beta$ & $-\cot\beta$ \\
Type II
& $\tan\beta$ & $\cot\beta$ & $\tan\beta$ \\
Lepton Specific
& $-\cot\beta$ & $\cot\beta$ & $\tan\beta$ \\
Flipped
& $\tan\beta$ & $\cot\beta$ & $-\cot\beta$ \\
\hline
\end{tabular}
\end{center}
\caption{The couplings $y^d_{A^0}$, $y^u_{A^0}$, and $y^\ell_{A^0}$  in the Yukawa interactions of $A^0$ in the four versions of the 2HDM with NFC.}
\label{2HDMAcoup}
\end{table}

The viable parameter space in a 2HDM must respect all theoretical and experimental constraints, which are listed below:
\begin{enumerate}
    \item Theoretical constraints:
    \begin{enumerate}
        \item[{(i)}] Vacuum stability of the 2HDM potential:\\
         The values of $\lambda_i$ are constrained by the requirement
        that the scalar potential a) breaks the electroweak symmetry $SU(2)_L\otimes U(1)_Y$
        to $U(1)_Q$, b) the scalar potential is bounded from below,  and c) the scalar potential  stays positive for arbitrarily large values of the scalar fields.  The constraints are:\\
         $\lambda _{1}> 0, \;\;\lambda _{2}> 0, \;\; \lambda _{3}+\lambda _{4}-\left|\lambda _{5} \right| + \sqrt{\lambda _{1} \lambda _{2}} \ge 0, \;\;\lambda _{3}+ \sqrt{\lambda _{1}\lambda _{2}}\ge 0$.\\
         From these conditions it be seen that $\lambda_1$ and $\lambda_2$ are positive definite, while $\lambda_3,  \lambda_4$ and $\lambda_5$ can have either sign. 
        \item[{(ii)}] Perturbativity:\\
         For calculational purposes it is required that the quartic couplings $\lambda _{i}$ do not take numerical values for which the perturbative expansion  ceases to converge.
        The couplings $\lambda _{i}$ remain perturbative up to the unification scale if 
        they satisfy the condition $\left| \lambda  _{i}\right|\leq 8\pi $.   
        \item[{(iii)}]  Unitarity: \\
        The  $2\to 2$ scattering processes ($s_1s_2\to s_3s_4$) involving only scalars (including Goldstone bosons) are mediated by scalar quartic couplings, which depend on the parameters of the scalar potential. 
        Tree-level unitarity constraints require that the eigenvalues of a scattering matrix of the amplitudes of $s_1s_2\to s_3s_4$ be less than the unitarity limit of $8\pi$, and this leads to further constraints on $\lambda_i$.
    \end{enumerate}
    \item Experimental constraints: 
     \begin{enumerate}
     \item[{(i)}] 
    Direct searches for Higgs bosons:\\
    The observation of the 125 GeV boson at the LHC and the non-observation of additional Higgs bosons at LEP, Tevatron and LHC rule out regions of the parameter space of a 2HDM.    
    In our numerical results these constraints are respected by using the publicly available codes HiggsBounds \cite{Bechtle:2020pkv} (which implements searches for additional Higgs bosons) and HiggsSignals \cite{Bechtle:2020uwn}
    (which implements the measurements of the 125 GeV boson). Any point in the 2HDM 
    parameter space that violates experimental limits/measurements concerning Higgs bosons is rejected. 
 \item[{(ii)}] Oblique parameters:\\
 The Higgs bosons in a 2HDM give contributions to the self-energies of the $W^\pm$ and $Z$ bosons.  The oblique parameters $S$, $T$ and $U$ \cite{Peskin:1990zt} describe the
 deviation from the SM prediction of $S=T=U=0$. The current best-fit values (not including the recent CDF measurement of $m_W$ \cite{CDF:2022hxs}) are \cite{ParticleDataGroup:2022pth}:
 \begin{equation}
 S=-0.01\pm 0.10,\;\; T=0.03\pm 0.12,\;\; U=0.02\pm 0.11 \,.
 \end{equation}
 
If $U=0$ is taken (which is approximately true in any 2HDM) then the experimental allowed ranges for $S$ and $T$ are narrowed to \cite{ParticleDataGroup:2022pth}:
 \begin{equation}
 S=0.00\pm 0.07,\;\; T=0.05\pm 0.06\,.
 \label{ST}
 \end{equation}
In our numerical results the theoretical constraints in 1(i), 1(ii), 1(iii) and the experimental constraints 2(ii) 
(using the ranges for $S$ and $T$ in eq.(\ref{ST})) are respected by using 2HDMC \cite{Eriksson:2009ws}. 
If the recent measurement of $m_W$ by the CDF collaboration \cite{CDF:2022hxs} is included in the world average for $m_W$
 then the central values of the $S$ and $T$ parameters in eq.(\ref{ST}) change significantly, and can be accommodated in a 2HDM
 by having sizeable mass splittings among the Higgs bosons. Recent studies have been carried out in 
\cite{Abouabid:2022lpg,Lee:2022gyf} in both NH and IH.

\item[{(iii)}] Flavour constraints:\\
The parameter space of a 2HDM is also constrained by flavour observables, especially the decays of $b$ quarks (confined inside $B$ mesons).
The main origin of such constraints is the fact that the charged Higgs boson $H^\pm$ contributes to processes that are mediated by a $W^\pm$, leading to constraints on the
parameters $m_{H^\pm}$ and $\tan\beta$. The flavour observable that is most constraining is the rare decay $b\to s\gamma$, although $H^\pm$ contributes to numerous
processes (e.g. $B\overline B$ mixing). There have been many studies of flavour constraints on the the parameter space of 2HDMs e.g.
\cite{Arbey:2017gmh,Atkinson:2022pcn,Cheung:2022ndq}.
In our numerical analysis we respect such flavour constraints by use of the publicly available code SuperIso \cite{Mahmoudi:2008tp}.
In the 2HDM (Type I), in which the couplings of $H^\pm$ to the fermions is proportional to $\cot\beta$, the constraint on $m_{H^\pm}$ is weaker with increasing $\tan\beta$.
The lowest value of  $\tan\beta$ we consider is $\tan\beta=3$, for which $m_{H^\pm}=140$ GeV is allowed (as can be seen in \cite{Arbey:2017gmh}).

 \end{enumerate}
  \end{enumerate}

\section{Phenomenology of $A^0$ at the LHC}
\noindent
In this section the formulae for the partial widths of $A^0$ are given and the previous studies of its BRs in the four types of 2HDM with NFC are summarised. The main 
production mechanisms for $A^0$ at the LHC are also discussed.
Emphasis will be given to the decay $A^0\to h^0Z^{(*)}$ for which there is a dependence on the mass of $h^0$ (we assume $m_{A^0}> m_{h^0}$).
In the NH one has $m_{h^0}=125$ GeV while in the IH the mass $m_{h^0}$ is a free parameter with $m_{h^0}<125$ GeV. Consequently,
the magnitude of BR$(A^0\to h^0Z^{(*)})$ in the parameter space of the 2HDM requires separate analyses in each of the two hierarchies.  Most previous studies of
the BRs of $A^0$ focus on the scenario of NH, with very few studies in the context of IH. These works will be summarised in this section.
 
\subsection{The Branching Ratios of $A^0$ in 2HDMs with NFC}
\noindent
We now present the explicit expressions for the partial decay widths of $A^0$ to a fermion ($f$) and an anti-fermion ($\overline f$) at tree-level . These generic expressions apply to all
four 2HDMs with NFC, with the model dependence arising in the $y^u_{A^0}$, $y^d_{A^0}$ and $y^\ell_{A^0}$ couplings that are displayed in Table \ref{2HDMAcoup}.
The partial widths $\Gamma(A^0\to f\overline f)$ are given by (e.g. see \cite{Djouadi:1995gv,Djouadi:2005gj,Branco:2011iw,Choi:2021nql}):
\begin{equation}
\Gamma(A^0\to u\overline u)=\frac{3G_F m_{A^0} m^2_u (y^u_{A^0})^2}{8\pi v^2}\lambda^{1/2}\left(\frac{m^2_u}{m^2_{A^0}}, \frac{m^2_u}{m^2_{A^0}} \right)\,,
\label{width_up}
\end{equation}
\begin{equation}
\Gamma(A^0\to d\overline d)=\frac{3G_F m_{A^0} m^2_d (y^d_{A^0})^2}{8\pi v^2}\lambda^{1/2}\left(\frac{m^2_{d}}{m^2_{A^0}}, \frac{m^2_d}{m^2_{A^0}} \right)\,,
\label{width_down}
\end{equation}
\begin{equation}
\Gamma(A^0\to \ell \overline \ell)=\frac{G_F m_{A^0} m^2_\ell (y^\ell_{A^0})^2}{8\pi v^2}\lambda^{1/2}\left(\frac{m^2_{\ell}}{m^2_{A^0}}, \frac{m^2_\ell}{m^2_{A^0}} \right)\,.
\label{width_tau}
\end{equation}
The phase space suppression factor is given by $\lambda(x,y)=(1-x-y)^2-4xy$.
For our main case of interest of $m_{A^0}>130$ GeV the factor $\lambda^{1/2}$ is essentially negligible for all fermions except the top quark (if $m_{A^0}>2m_t$).
In the above expressions the running quark masses $m_u$ and $m_d$ are evaluated at the energy
scale ($Q$) of $m_{A^0}$, and this encompasses the bulk of the QCD corrections.
There are also QCD vertex corrections to the decays to quarks which have the effect of multiplying the above partial widths by an overall factor.
To order $\alpha_s$ this factor is given by $(1+17\alpha_s/(3\pi))$ and higher-order vertex corrections have been calculated  \cite{Djouadi:2005gj}.

The partial width for the decay to two gluons ($A^0\to gg$) at leading order is mediated by triangle loops of fermions. The dominant contribution comes from i) the triangle diagram with $t$-quarks, which is proportional to $(y^t_{A^0})^2$, and ii) the 
triangle diagram with 
$b$-quarks, which is proportional to $(y^b_{A^0})^2$.
The explicit formula for $\Gamma(A^0\to gg)$ can be found in \cite{Djouadi:1995gv,Djouadi:2005gj,Spira:1995rr}.
There is the also the decay $A^0\to \gamma\gamma$, which is mediated by triangle loops of $f$, $W^\pm$ and $H^\pm$. However, $\Gamma(A\to \gamma\gamma)$
is much smaller than $\Gamma(A^0\to gg)$ because the former has a factor of $\alpha^2$ while the latter has a factor of $\alpha^2_s$. 
The decays $A^0\to W^+W^-$ and $A^0\to ZZ$ are absent at tree-level in the (CP-conserving) 2HDM. These decays are generated at higher orders
but have much smaller BRs \cite{Bernreuther:2010uw,Arhrib:2018pdi} than some of the tree-level decays and will be neglected in our study.

Finally, we consider the decays of $A^0$ to another Higgs boson and to a vector boson, which can be dominant. These
interactions originate from the kinetic term in the Lagrangian and do not involve the Yukawa couplings. 
The partial width for $A^0\to h^0 Z$ (i.e. a two-body decay with on-shell $Z$) is given by:
\begin{equation}
\Gamma(A^0\to h^0 Z)=\frac{m^3_{A^0}\cos^2(\beta-\alpha)}{v^2}\lambda^{3/2}\left(\frac{m^2_{h^0}}{m^2_{A^0}}, \frac{m^2_Z}{m^2_{A^0}} \right)\,.
\label{width_AhZ}
\end{equation}
The partial width $\Gamma(A^0\to h^0 Z^*)$ (i.e.  a three-body decay with off-shell $Z^*\to f\overline f$)
is also proportional to $\cos^2(\beta-\alpha)$ and 
involves an integration over the momenta of $f\overline f$. Its explicit expression is given in  \cite{Djouadi:1995gv,Moretti:1994ds,Aiko:2022gmz}.
The partial width $\Gamma(A^0 \to H^0Z)$ has the same form as eq.~(\ref{width_AhZ}), but with $m_{h^0}$ replaced by $m_{H^0}$ and
$\cos^2(\beta-\alpha)$ replaced by $\sin^2(\beta-\alpha)$.
We do not consider the decay channel $A^0\to H^\pm W^\mp$ as we shall be taking $m_{A^0}=m_{H^\pm}$.

We now briefly review previous studies of the decay  $A^0\to h^0 Z^{(*)}$, which were first performed in the context of the Minimal Supersymmetric Standard Model (MSSM).
 The scalar potential of the MSSM takes the form of the scalar potential of the 2HDM but with fewer free parameters in it and necessarily Type II Yukawa couplings.
In the MSSM $m_{h^0}$ has an upper bound of around 130 GeV, in which $m_{h^0}=125$ GeV can be accommodated with large SUSY corrections to the tree-level scalar potential.
The value of $\sin^2(\beta-\alpha)$ rapidly approaches 1 as $m_{A^0}$ increases above 100 GeV and this is in contrast to a non-SUSY 2HDM for which 
$\sin^2(\beta-\alpha)$ could differ substantially from 1 for $m_{A^0}>100$ GeV. Early studies of BR($A^0\to h^0 Z$) in the MSSM and its detection prospects at the LHC can be found in \cite{Gunion:1991cw, Baer:1992uu,Abdullin:1996as}.
The first calculation of $\Gamma(A^0\to h^0 Z^*)$ was carried out in \cite{Djouadi:1995gv,Moretti:1994ds}, but this three-body decay has limited importance in the MSSM due its Type II structure and the fact that $\cos(\beta-\alpha)$
rapidly tends to zero as $m_{A^0}$ increases.
The BRs of $A^0$ in the MSSM are summarised in \cite{Djouadi:2005gj}. For low $\tan\beta$ (e.g. $ \tan\beta=3$), BR$(A^0\to h^0 Z^{(*)})$ can be of the order of 10\% or more in the region $200\,{\rm GeV} < m_{A^0} < 300$ GeV when the
two-body decay is open and before $A^0\to t\overline t$ becomes dominant for heavier $m_{A^0}$.

In the context of non-supersymmetric 2HDMs with NFC (on which we focus) an early study of the on-shell decay $A^0\to h^0 Z$ (Type I and Type II only) was carried out in \cite{Kominis:1994fa}, taking 
several values of $\sin^2(\beta-\alpha)$ in the range $0\to 1$ and $m_{h^0}=100$ GeV. It was shown that this decay channel for $A^0$ can have the largest BR, and detection prospects at the LHC in the channel
$A^0\to h^0 Z\to \gamma\gamma \ell^+\ell^-$ were studied. 
The three-body decay $A^0\to h^0Z^*$ in non-supersymmetric 2HDMs (Type I and Lepton Specific) with NFC were first studied in the context of LEP2 in \cite{Akeroyd:1998dt}. It was pointed out
that BR($A^0\to h^0Z^*)$ can be dominant in Type I as $\tan\beta$ increases because $\Gamma(A^0\to f\overline f)$ decreases $\propto \cot^2\beta$. This in contrast to the case in the MSSM where BR($A^0\to h^0Z^*)$ is always small.
In \cite{Akeroyd:1998dt}, BR($A^0\to h^0Z^*)$ was studied as a function of $\tan\beta$ in the 2HDM (Type I) for $m_{A^0}=80$ GeV, 100 GeV and 120 GeV, with $m_{h^0}=40$ GeV and $\cos^2(\beta-\alpha)=1$. This is the IH scenario but at that time $m_{H^0}$ was not known.

Studies of the BRs of $A^0$ in the four versions of the 2HDM with NFC were given in \cite{Aoki:2009ha} for $m_{A^0}=150$ GeV without including $A^0\to h^0 Z^{(*)}$ ($m_{h^0}$ or $m_{H^0}=125$ GeV was not known at the time).
Recent works \cite{Aiko:2022gmz} have presented the BRs of $A^0$ including  $A^0\to h^0 Z^{(*)}$ in the scenario of NH ($m_{h^0}=125$ GeV) with
$\sin^2(\beta-\alpha)\approx 1$ and these results will be summarised below.
Electroweak corrections to $\Gamma(A^0\to h^0 Z)$ were also calculated for the first time in \cite{Aiko:2022gmz} and are of the order of $10\%$.

\begin{table}[h]
\begin{center}
\begin{tabular}{|c||c|c|}
\hline
2HDM Parameter & Normal Hierarchy (NH) &  Inverted Hierarchy (IH)  \\ \hline
$m_{h^0}$ & 125 GeV &  $10\,{\rm GeV}< m_{h^0} < 100$ GeV   \\ \hline
$m_{H^0}$ & $ 300$  GeV & 125 GeV  \\ \hline
$m_{A^0}$ &  130 GeV $\le m_{A^0} \le 400$ GeV  & 130 GeV $\le m_{A^0} \le 400$ GeV \\ \hline
$m_{H^\pm}$ & $m_{H^0}$ & $m_{A^0}$ \\ \hline
$\cos(\beta-\alpha)$ &  $0 \le |\cos(\beta-\alpha)|< 0.1$ &   $0.9 < |\cos(\beta-\alpha)|< 1$   \\  \hline
$\tan\beta$ & $2.9\le \tan\beta \le 5.2$   & $2.9\le \tan\beta\le 5.2$  \\  \hline
$m^2_{12}$ & $560 \,{\rm GeV}^2 \le m^2_{12} \le  1670\,{\rm GeV}^2$   &  $560 \,{\rm GeV}^2 \le m^2_{12} \le  1670 \,{\rm \,GeV}^2$ \\  \hline
\end{tabular}
\end{center}
\caption{2HDM parameter ranges in NH ($m_{h^0}=125$ GeV) and IH ($m_{H^0}=125$ GeV) that will be considered in this work. Some attention will also be given to the region
$80\,{\rm GeV} < m_{A^0}+m_{h^0} < {\rm 110 GeV}$.}
\label{2HDM-Param}.
\end{table}

The ranges of the five parameters $m_{h^0}$, $m_{H^0}$, $m_{A^0}$, $\tan\beta$ and $\cos(\beta-\alpha)$ that will be considered in this work are given in
Table \ref{2HDM-Param}. The parameter $\tan\beta$ only takes positive values, while $\cos(\beta-\alpha)$ can take positive or negative values.
In the case of NH one has (by definition)  $m_{h^0}=125$ GeV and so necessarily $m_{H^0}>125$ GeV. 
The discovered 125 GeV boson has been measured by the LHC experiments to have SM-like Higgs boson couplings within experimental error, and in the context of a 2HDM with NH
 the parameter $|\cos(\beta-\alpha)|$ is thus constrained to be (approximately) less than 0.1.
The exact constraint on $|\cos(\beta-\alpha)|$ has a dependence on $\tan\beta$, as well as a dependence on which 2HDM is being considered e.g. in the 2HDM (Type II), $|\cos(\beta-\alpha)|= 0.1$
is only possible
for $1 < \tan\beta < 2$, while in the 2HDM (Type I), $|\cos(\beta-\alpha)|$ can reach a value of 0.25 for $3 < \tan\beta < 5$, with $|\cos(\beta-\alpha)|=0.1$ being  possible up to large values of $\tan\beta$. 
In the 2HDM (Type II) and 2HDM (Flipped) there is a very small region (disconnected from the aforementioned region) of $\cos(\beta-\alpha)\approx 0.25$ for $\tan\beta\approx 10$. This latter region is called the
"wrong-sign" Yukawa coupling region and will be discussed in more detail in Sec.~IVC.
The LHC measurements also constrain the sign of $\cos(\beta-\alpha)$ and for a given value of $\tan\beta$ the constraint on 
 $\cos(\beta-\alpha)$ is in general different for its positive and negative values. Since the coupling $A^0 h^0 Z$ is proportional to $\cos(\beta-\alpha)$, in NH the
 decay channel $A^0\to h^0 Z$ has a suppression factor of $|\cos(\beta-\alpha)|^2\approx 0.01$.
 Despite this suppression factor, BR$(A^0\to h^0 Z$) can still be sizeable (or dominant) in regions of parameter space of the four 2HDMs.
 In \cite{Aiko:2020ksl}, the BRs of $A^0$ were shown for $\sin(\beta-\alpha)=0.995$ and $m_{A^0}=200$ GeV, for which $A^0\to h^0Z^*$ is a three-body decay. In the 2HDM (Type I)
 $A^0\to h^0Z^*$ has the largest BR for $\tan\beta>20$, but in the other three models BR$(A^0\to h^0Z^*)<1\%$. In \cite{Aiko:2022gmz} the 2HDM parameters were changed to
 $m_{A^0}=300$ GeV (for which $A^0\to h^0Z$ is a two-body decay) and the range
 $0 < |\cos(\beta-\alpha)|<0.1$ was considered. It was shown that $A^0\to h^0Z$ has the largest BR in all four models for $|\cos(\beta-\alpha)|$ closer to its upper limit of 0.1, with 
 the 2HDM (Type I) having the largest parameter space for $A^0\to h^0Z$ being the dominant decay. 
  
 In the case of the IH one has $m_{H^0}=125$ GeV and so necessarily $m_{h^0}< 125$ GeV. The above constraints on $\cos(\beta-\alpha)$ now apply to 
 $\sin(\beta-\alpha)$, and so $0.9 < |\cos(\beta-\alpha)|< 1$. Hence the decay $A^0\to h^0 Z$ has very little suppression from the coupling $A^0 h^0 Z$, in contrast to the case
 of NH. Moreover, since $m_{h^0}<125$ GeV the decay $A^0\to h^0Z$ can proceed via an on-shell $Z$ for lighter values of $m_{A^0}$ than in the case of NH
 i.e. $m_{A^0} > 216$ GeV is required for on-shell $A^0\to h^0Z$ if $m_{h^0}=125$ GeV, but for $m_{h^0}=90$ GeV (say) then the on-shell decay $A^0\to h^0Z$ is open for $m_{A^0}> 180$ GeV.
 Moreover, off-shell decays $A^0\to h^0Z^*$ can also be dominant in the 2HDM (Model I) over a large region of parameter space of the model. The BRs of $A^0$ in the scenario of IH
 will be studied in detail in section V. 
In the case of IH the mass $m_{h^0}$ ($< 125$ GeV) is 
 an unknown parameter and the BRs of $h^0$ will be different (in general) to those of the SM-like 125 GeV Higgs boson.
 Previous studies of BR($A^0\to h^0Z^{(*)}$) in the 2HDM (Type I) in IH are rare, and include an early study in \cite{Akeroyd:1998dt} (as mentioned above, for 80 GeV $< m_{A^0}< 120$ GeV) and more recently in 
 \cite{Abouabid:2022lpg} in which BR($A^0\to h^0Z^{(*)})$ was shown as a scatter plot with 60 GeV $< m_{A^0}< 600$ GeV. Another recent work \cite{Moretti:2022fot} also makes use of the potentially large BR($A^0\to h^0Z^{(*)})$ 
 and this will be described in the next paragraph.

The parameter space of $m_{h^0}+m_{A^0}<200$ GeV is strongly constrained by the fact that there was no signal
in the channel $e^+e^-\to Z^* \to A^0 h^0\to b\overline b b\overline b$ at LEP2. In a 2HDM (Type I) in the IH scenario one has $\cos(\beta-\alpha)\approx 1$, which
maximises the coupling $ZA^0h^0$ and suggests $m_{h^0}+m_{A^0}>200$ GeV from the above channel. However, recently in Ref.~\cite{Moretti:2022fot} it has been shown that
 $m_{h^0}+m_{A^0} < 200$ GeV is still possible in IH provided that BR$(A^0\to b\overline b$) is suppressed due to a large BR$(A^0\to h^0 Z^*)$. 
 In Ref.~\cite{Moretti:2022fot} several benchmark points (which satisfy all current constraints) were listed with $80\,{\rm GeV} < m_{h^0}+m_{A^0} < 110$ GeV.
 In this parameter space BR$(A^0\to h^0 Z^*)$ can be large for the same reasons outlined in \cite{Akeroyd:1998dt}, although this latter work only showed results for $m_{h^0}+m_{A^0}>120$ GeV. 
 All benchmark points have the mass hierarchy $m_{H^0}(=125\,{\rm GeV}) > m_{A^0}>m_{h^0}$ and a light charged Higgs boson in the range $100\,{\rm GeV}< m_{H^\pm} < 160$ GeV. 
 It was suggested in Ref.~\cite{Moretti:2022fot} that this parameter space of  $80\,{\rm GeV} < m_{h^0}+m_{A^0} < 110$ GeV could be probed via the mechanism
 $gg\to H^0\to A^0Z^*\to h^0Z^*Z^*$, with subsequent decays $h\to b\overline b$ and $Z^*Z^*\to jj\mu^+\mu^-$, and a simulation of its detection prospects was carried out.
 It was shown that $\sigma(gg\to H^0\to A^0Z^*\to h^0Z^*Z^*)$ can reach 0.01 pb, with BR($H^0\to A^0 Z^*)$ having a maximum value of $0.2\%$ and being a
 significant suppression factor. A number of benchmark points have a statistical significance of $2\sigma$ to $3\sigma$ (a few reaching $4\sigma$) for an integrated luminosity
of 300 fb$^{-1}$, and roughly scaling by a factor of 3 with 3000 fb$^{-1}$ at the HL-LHC. 
The channel to be studied in this work, $gg\to A^0\to h^0 Z^*$, would also be a probe of this scenario of $m_{h^0}+m_{A^0}< 200$ GeV, although our main focus
will be on the region $m_{A^0}+m_{h^0}>200$ GeV. We shall compare $\sigma(gg\to A^0\to h^0 Z^*)$ with $\sigma(gg\to H^0\to A^0Z^*\to h^0Z^*Z^*)$ for some of the benchmark points in \cite{Moretti:2022fot}.

\subsection{Production mechanisms for $A^0$ at the LHC}
\noindent
At the LHC the main production processes for $A^0$ are \cite{Djouadi:2005gj,Spira:1995rr,Bagnaschi:2022dqz}:\\
i)  $gg\to A^0$ (gluon-gluon fusion), which proceeds via a top-quark loop and a bottom-quark loop, and thus involves the Yukawa couplings for the vertices $A^0 t\overline t$ and $A^0 b\overline b$. \\
 ii) $gg\to A^0 b\overline b$ 
(associated production with $b$ quarks), which depends on the Yukawa coupling for the vertex $A^0 b\overline b$.\\

Both mechanisms involve the couplings of $A^0$ to fermions
and hence their respective cross sections depend on which 2HDM is under consideration (see Table~\ref{2HDMAcoup}). For $gg\to A^0$ the top-quark loop is dominant in all four 2HDMs
for lower values of $\tan\beta$ (e.g. $\tan\beta<5$). For larger values of $\tan\beta$ (e.g. $\tan\beta>5$) 
the top-quark loop is still dominant in the 2HDMs Type I and Lepton Specific, but $\sigma(gg\to A^0)$
decreases with increasing $\tan\beta$ because the top-quark and bottom-quark Yukawa couplings are both proportional to $\cot\beta$.
In contrast,  in the Type II and Flipped 2HDMs the bottom-quark loop becomes the dominant contribution to $\sigma(gg\to A^0)$ for larger
values of $\tan\beta$ because the bottom-quark Yukawa coupling is proportional to $\tan\beta$. Hence $\sigma(gg\to A^0)$ increases with increasing $\tan\beta$ after reaching a minimum at around $\tan\beta\approx 7$.
The production mechanism $gg\to A^0 b\overline b$ does not involve the top-quark Yukawa coupling and is only relevant in the Type II and Flipped 2HDMs for larger values of $\tan\beta$, 
for which it has a larger cross section than $\sigma(gg\to A^0)$. In the Type I and Lepton Specific 2HDMs one always has $\sigma(gg\to A^0 b\overline b)<\sigma(gg\to A^0)$. 
The numerical values of both cross sections in the plane $[m_{A^0}, \tan\beta]$ are presented in \cite{Aiko:2020ksl}.  For $m_{A^0}=200$ GeV 
both cross sections can be greater than 100 pb, depending on the 2HDM under study and the value of $\tan\beta$.

\section{Searches for $A^0\to h^0 Z$ at the LHC}
\noindent
The decay $A^0\to h^0 Z$ has been searched for
at the LHC by the ATLAS and CMS  collaborations assuming the case of NH  (i.e. $m_{h^0}=125$ GeV) and 
an on-shell $Z$ boson. These searches will be summarised in this section. No search has yet been carried out for $A^0\to h^0 Z$ in IH.
Current LHC searches for $A^0\to h^0 Z$ (to be described below) assume that $m_{h^0}=125$ GeV
 and $m_{A^0} \ge 225$ GeV. 
In this work we will focus on the mass range 130 GeV $\le m_{h^0}+m_{A^0}  \le 400$ GeV in the context of the IH scenario ($m_{h^0} < 125$ GeV and $m_{H^0}$=125 GeV).
Some discussion will also be given to the case of 80 GeV $\le m_{h^0}+m_{A^0}  \le 110$ GeV.
 
 \begin{table}[h]
\begin{center}
\begin{tabular}{|c||c|c|}
\hline
$\sqrt s \;($integrated luminosity)& ATLAS &  CMS  \\ \hline
8 TeV (20 fb$^{-1}$)
&  $b\overline b \ell\overline \ell$/$b\overline b\nu\overline \nu$ \cite{ATLAS:2015kpj}, $\tau\overline\tau\ell\overline\ell$ \cite{ATLAS:2015kpj}
&   $b\overline b \ell\overline \ell$ \cite{CMS:2015flt}, $\tau\overline\tau\ell\overline\ell$ \cite{CMS:2015uzk}\\
13 TeV (35.9 fb$^{-1}$)
&  $b\overline b \ell\overline \ell$/$b\overline b\nu\overline \nu$ \cite{ATLAS:2017xel} & $b\overline b \ell\overline \ell$/$b\overline b\nu\overline \nu$ \cite{CMS:2019qcx}, $\tau\overline\tau\ell\overline\ell$ \cite{CMS:2019kca}\\
13 TeV (139 fb$^{-1}$)
& $b\overline b \ell\overline \ell$/$b\overline b\nu\overline \nu$ \cite{ATLAS:2020pgp}  &  \\
\hline
\end{tabular}
\end{center}
\caption{Searches for $A^0\to h^0 Z$ at the LHC, using $gg\to A^0$ and $gg\to A^0 b\overline b$ as the production mechanism, and taking $m_{h^0}=125$ GeV. The integrated
luminosities used for the searches are given in brackets next to the collider energy $\sqrt s$. The four-fermion signature $b\overline b \ell\overline \ell$ means that
$h^0\to b\overline b$ and $Z\to \ell\overline \ell$, where $\ell$ denotes $e$ or $\mu$ (i.e. the decays of $h^0$ are given first).}
\label{LHC_search}
\end{table}
The searches for $A^0\to h^0 Z$ at the LHC, using $gg\to A^0$ and $gg\to A^0 b\overline b$ as the production mechanisms, are summarised in Table~\ref{LHC_search}.
Two decays channels of $h^0$ are targeted, namely $h^0\to b\overline b$ and $h^0\to \tau\overline \tau$.
In both searches $A^0$ is assumed to be produced via $gg\to A^0$ and $gg\to A^0 b\overline b$ with subsequent decay 
via the channel $A^0\to h^0 Z$ in which $Z$ is on-shell. Hence the searches probe $m_{A^0}> m_{h^0}+m_Z$ ($\approx 216$ GeV), and limits are shown for $m_{A^0}>225$ GeV only.
 In the context of the NH ($m_{h^0}=125$ GeV) the magnitudes of
these BRs of $h^0$ to fermions are given by the measurements of the BRs of the 125 GeV boson, and thus BR($h^0\to b\overline b)\approx 57\%$ and BR($h^0\to \tau\overline \tau)\approx 6\%$ (i.e. roughly the same as
the BRs of the SM Higgs boson). In the IH
case on which we focus, these BRs of $h^0$ will be in general different from those in the case of the NH,  with a dependence on (the unknown) $m_{h^0}$.

\subsection{LHC search for $A^0\to h^0Z\to b\overline b \ell^+\ell^-$}
\noindent
We now discuss the search by CMS for the signatures $b\overline b \ell\overline \ell$/$b\overline b\nu\overline \nu$ \cite{CMS:2019qcx} with
$\sqrt s=13$ TeV and 35.9 fb$^{-1}$ of integrated luminosity. In both searches $A^0$ is assumed to be produced via $gg\to A^0$ and $gg\to A^0 b\overline b$ with subsequent decay 
via the channel $A^0\to h^0 Z$ in which $Z$ is on-shell.
In \cite{CMS:2019qcx}, which only targets the decay channel $h^0\to b\overline b$ ($m_{h^0}=125$ GeV), separate searches in each production channel are carried out for:\\
i) the decays $Z\to e^+e^-$ and $Z\to \mu^+\mu^-$ (collectively referred to as $Z\to \ell\overline \ell$), leading to
the signature $b\overline b \ell\overline \ell$.\\
ii) the decay $Z\to \nu\overline\nu$, leading to the signature $b\overline b \nu\overline \nu$.\\
In each of i) and ii) above, the signal is separated into categories with 1 $b$ quark, 2 $b$ quarks and 3 $b$ quarks.
In what follows we will focus on the signature $b\overline b \ell\overline \ell$ because the $Z\to \nu\overline\nu$ signature has no sensitivity for $m_{A^0}<500$ GeV, and is
is only competitive with the $b\overline b \ell\overline \ell$ signature for $m_{A^0}>700$ GeV.
For the $b\overline b \ell\overline \ell$ signature in i) above, the selection efficiencies are similar for the $gg\to A^0$ and $gg\to A^0 b\overline b$ production mechanisms in the 1 $b$-quark and 2 $b$-quark categories, and these
efficiencies increase slightly with
 increasing $m_{A^0}$. In the 3 $b$-quark category, the selection efficiency for $gg\to A^0 b\overline b$ is considerably larger (due to the presence of more $b$ quarks in the signal) 
 than that for $gg\to A^0$, being almost an order of magnitude greater for $m_{A^0}<300$ GeV.
 The SM backgrounds to the $b\overline b \ell\overline \ell$ (and $b\overline b \nu\overline \nu$) signatures are largest for the 1 $b$-quark category and smallest for the 3 $b$-quark category. 

The invariant masses of $b\overline b \ell\overline \ell$ events which pass all the selection cuts are displayed starting from 225 GeV.
A clear signal for $A^0\to h^0 Z$ would appear as a peak centred on $m_{A^0}$ above the background.
For the background (which mainly arises from processes $Z$+jets, $Z+b$, $Z+ b\overline b$, $t\overline t$) the invariant mass distribution of $b\overline b \ell\overline \ell$ events
rises up to a peak at around 250 GeV before falling in all three $b$-quark categories. For the $b\overline b \nu\overline \nu$ signature in ii) above, in both production modes the selection efficiencies in a particular $b$-quark category are much smaller than those for $b\overline b \ell\overline \ell$ in the same $b$-quark
category for $m_{A^0} < 500$ GeV, but become similar in magnitude for $m_{A^0}>600$ GeV. For the background, the transverse mass of $b\overline b \nu\overline \nu$ 
 (starting from 500 GeV) decreases in all $b$-quark categories.
 
 In the NH scenario one has $m_{h^0}=125$ GeV and hence the invariant mass distribution of the $b\overline b$ pair originating from $h^0$ (i.e. the signal) would be centred on 125 GeV. This would
 not be true for the background, and to exploit this fact an invariant mass cut of $100\;{\rm GeV}< m_{b\overline b}< 140$ GeV is imposed in the CMS search in \cite{CMS:2019qcx}.
 This cut preserves most of the signal while reducing the backgrounds. The events with $m_{b\overline b} <100\;{\rm GeV}$ and $m_{b\overline b}> 140$ GeV are 
put into the sidebands.  However, in the IH scenario (for which $m_{b\overline b}$ would peak at a lower value than 125 GeV) the above cut on $m_{b\overline b}$ would be moving potential signal events to the sidebands.
The CMS search also requires a cut of $70\,{\rm GeV} < m_{\ell\overline \ell} <  110\,{\rm GeV}$
on the invariant mass of the leptons originating from $Z$. This cut captures most of the
leptons originating from the decay of an on-shell $Z$, but would not be as effective for an off-shell $Z^*$ (e.g. in the case of BR($A^0\to h^0 Z^*$) being large in IH).

The expected limits on $\sigma(gg\to A^0)\times {\rm BR}(A^0\to h^0 Z\to b\overline b \ell\overline \ell)$ are found to be 45 fb for $m_{A^0}=225$ GeV and falling to
10 fb for $m_{A^0}=400$ GeV. 
The lack of any statistically significant signal in the search in \cite{CMS:2019qcx} allows constraints to be obtained on the 2HDM parameter space of $[\cos(\beta-\alpha), m_{A^0}, \tan\beta]$.
Taking $\cos(\beta-\alpha)=0.1$ (which is motivated from the experimental fact that $h^0$ has SM-like couplings, $\sin^2(\beta-\alpha)\approx 1$) limits are shown in the plane $[m_{A^0}, \tan\beta]$.
In the 2HDM (Type I) the dominant production process for all $\tan\beta$ is $gg\to A^0$, and the constraint on $\tan\beta$ strengthens from around $\tan\beta>4$ to
$\tan\beta>10$ as $m_{A^0}$ increases from 225 GeV to 350 GeV. For $m_{A^0}>350$ GeV the presence of the decay channel $A^0\to t\overline t$ reduces BR$(A^0\to h^0Z)$ and leads to a weakening
of the bound to $\tan\beta>1$ for $m_{A^0}>400$ GeV. Very similar limits are obtained in the Lepton Specific 2HDM. In the 2HDMs (Type II and Flipped) the limit on low values of $\tan\beta$ is weaker, being
 $\tan\beta>2$ to $\tan\beta>4$ as $m_{A^0}$ increases from 225 GeV to 350 GeV. However, in these latter two models the bottom-loop contribution to the production process $gg\to A^0$ and
 the process $gg\to A^0b\overline b$ are both enhanced at large $\tan\beta$, and this leads to limits of $\tan\beta<20$ for $m_{A^0}>450$ GeV.
 
 The searches for the signature $b\overline b \ell\overline \ell$/$b\overline b\nu\overline \nu$\ by the ATLAS collaboration in \cite{ATLAS:2017xel} 
 and \cite{ATLAS:2020pgp} have similar strategies and derive comparable limits on the parameter space of the 2HDM. The search with 36.1 fb${^{-1}}$  \cite{ATLAS:2017xel} 
 presents results for $m_{A^0}>220$ GeV while the search with 139 fb${^{-1}}$ \cite{ATLAS:2020pgp}
 presents results for $m_{A^0}>280$ GeV.

\subsection{LHC search for $A^0\to h^0Z\to \tau\overline \tau \ell^+\ell^-$}
\noindent
We now discuss the search by CMS for the signatures  $\tau\overline\tau\ell^+\ell^-$ \cite{CMS:2019kca} with
$\sqrt s=13$ TeV and 35.9 fb$^{-1}$ of integrated luminosity. This signature requires the decay $h^0\to \tau^+\tau^-$ , which has a 
BR of around 6\% and is almost 10 times smaller than BR$(h^0\to b\overline b)=57\%$. Consequently, the limits on the 2HDM
parameter space from the $\tau\overline \tau \ell^+\ell^-$ signature are somewhat weaker than those from the search for
$b\overline b \ell\overline \ell$. 

A $\tau$ lepton can decay hadronically (i.e. to hadrons accompanied by missing energy in the form of neutrinos) or leptonically (to an $e^\pm$ or $\mu^\pm$, with missing energy).
Four signatures from the decay $h^0\to \tau^+\tau^-$are considered, where $\tau_h$ denotes a $\tau^\pm$ that decays
hadronically: $e\tau_h$, $\mu\tau_h$,$\tau_h\tau_h$, $e\mu$. The $Z$ boson is taken to decay to $e^+e^-$ or $\mu^+\mu^-$, giving rise to 8 different channels
for the signature $\tau\overline\tau\ell^+\ell^-$. All 8 channels are combined when deriving the limits on 
$\sigma(gg\to A^0)\times {\rm BR}(A^0\to h^0 Z\to \tau\overline \tau \ell\overline \ell)$. 

The irreducible backgrounds are $ZZ(\to 4\ell)$, $t\overline t Z$, $WWZ$, $WZZ$ and $ZZZ$. The reconstructed pseudoscalar mass $m_{A^0}$,
denoted by $m^c_{\ell\ell\tau\tau}$, is used as the discriminant between the signal and the background. The simplest reconstructed mass (denoted by $m^{vis}_{\ell\ell\tau\tau}$) 
is obtained from the visible decay products only, but $m^c_{\ell\ell\tau\tau}$ significantly improves the mass resolution by accounting for the missing energy in the 
decays of $\tau^\pm$ and also using $m_{h^0}=125$ GeV (which is true in NH only) as input in the fitting procedure.

The expected limits on $\sigma(gg\to A^0)\times {\rm BR}(A^0\to h^0 Z\to \tau\overline \tau \ell\overline \ell)$ are found to be 13 fb for $m_{A^0}=220$ GeV and falling to
5 fb for $m_{A^0}=400$ GeV. These limits are somewhat stronger than those for the  $b\overline b \ell\overline \ell$ signature (where the limits are
45 fb for $m_{A^0}=220$ GeV and 10 fb for $m_{A^0}=400$ GeV). However, due to BR$(h^0\to \tau^+\tau^-$)/BR$(h^0\to b\overline b)\approx 0.1$
the limits on the 2HDM parameter space (which arise from $\sigma(gg\to A^0)\times {\rm BR}(A^0\to h^0 Z$) only) are stronger from the $b\overline b \ell\overline \ell$ signature.

\subsection{Case of $A^0\to h^0Z^{*}$ in NH and for the 2HDM (Type II)}
\noindent
None of the above searches considered the case of the off-shell decay $A^0\to h^0 Z^*$.
All searches targeted the mass region of $m_{A^0} > m_{h^0}+m_{Z}$ so that the $Z$ boson in the decay $A^0\to h^0 Z$
is always on-shell. A study in \cite{Accomando:2020vbo} considered the detection prospects
in the region $m_{A^0}<225$ GeV in NH and the 2HDM (Type II). Although BR($A^0\to h^0 Z^{(*)})$ is decreasing as
$m_{A^0}$ is lowered below 225 GeV, the background is also decreasing and is rather small for $m_{A^0}<210$ GeV. 
Three benchmark points were
chosen, with values of $m_{A^0}$, $\cos(\beta-\alpha)$ and $\tan\beta$ as follows:\\
 i)   $m_{A^0}=190$ GeV,   $\cos(\beta-\alpha)=0.36 $, $\tan\beta=4.9$. \\
 ii)  $m_{A^0}=200$ GeV,   $\cos(\beta-\alpha)=0.28$, $\tan\beta=6.4$. \\
iii)  $m_{A^0}=210$ GeV,   $\cos(\beta-\alpha)=0.26$, $\tan\beta=6.9$. \\
These benchmark points all correspond to the scenario of "wrong sign" down-type Yukawa coupling. This is a limit in which the down-type Yukawa couplings for $h^0$ in NH
in the 2HDM (Type II) are equal in magnitude to their values in the SM but with opposite sign. The wrong-sign limit is obtained for the choice of $\alpha+\beta=\pi/2$, and can be displayed
as all points on a hyperbola in the plane of $[\cos(\beta-\alpha), \tan\beta]$ going from points of large $\tan\beta$ ($\beta\approx \pi/2$) and  $\cos(\beta-\alpha)\approx 0$ (i.e. $\alpha\approx 0$, so that
$\alpha+\beta=\pi/2$) to points of small $\tan\beta$ ($\beta\approx \pi/4$) and $\cos(\beta-\alpha)\approx 1$ ($\alpha\approx \pi/4$, so that $\alpha+\beta=\pi/2$). The wrong-sign scenario
allows larger values of $\cos(\beta-\alpha)$ than in the alignment scenario, the latter being defined by $\beta-\alpha=\pi/2$ and consequently $\cos(\beta-\alpha)$ is close to zero.  Due to the fact that $\Gamma(A^0\to h^0 Z^*)\propto \cos^2(\beta-\alpha)$,
in the wrong-sign scenario BR$(A^0\to h^0 Z^*)$ can be larger than in the alignment scenario.

 The latest LHC measurements of the couplings of $h^0$ ($m_{h^0}=125$ GeV) now restrict the wrong-sign region in the  2HDM (Type II) to
points on the hyperbola for $\tan\beta>7$ and $|\cos(\beta-\alpha)|<0.3$ and so the above benchmark points are now either excluded or just allowed by the current experimental measurements. 
It was shown in \cite{Accomando:2020vbo} with a parton-level simulation that the detection prospects for $A^0\to h^0 Z^*$ at the LHC with 1000 fb$^{-1}$ were reasonable in each of the three benchmark points,
although a more detailed simulation would be needed to account for effects beyond the parton-level and at the level of the LHC detectors. We emphasise that the study in  \cite{Accomando:2020vbo} was not carried out
in the context of IH. In section V we shall consider $m_{A^0}<225$ GeV and $A^0\to h^0 Z^{(*)}$ in the IH scenario in the 2HDM (Type I) with NFC.

\section{Results}
\begin{figure}
    \centering
    \includegraphics[width=17cm]{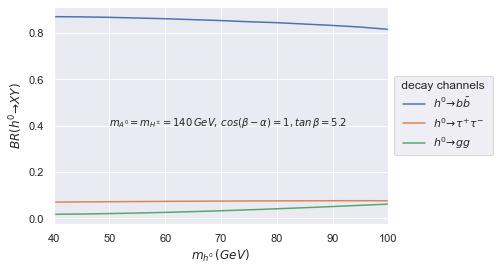}
    \caption{ The BRs of $h^0$ in the 2HDM (Type I) as a function of $m_{h^0}$ in IH ($m_{H^0}=125$ GeV) with $\cos(\beta -\alpha )=1$, $\tan\beta=5.2$ and $m_{A^0}=m_{H^\pm}=140$ GeV.}
    \label{fig:h0BR}
    \end{figure}

\begin{figure}
    \centering
    \includegraphics [width=17cm]{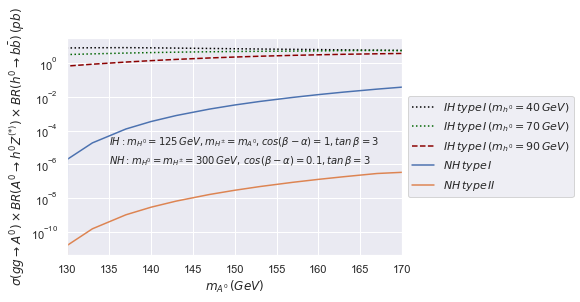}
    \caption{The signal cross section $\sigma(gg\to A^0)\times {\rm BR}(A^0\to h^0Z^{(*)})\times {\rm BR}(h^0\to b\overline b)$ as a function of $m_{A^0}$  
for NH with Type I, NH with Type II and for three choices of $m_{h^0}$ in IH with Type I.  The values of the input parameters are
 displayed on the figure.  }
    \label{fig:sigmaBR170}
\end{figure}

\noindent
In this section we show our results for the signal cross section, which is given by the following product:
\begin{equation}
\sigma(gg\to A^0)\times {\rm BR}(A^0\to h^0Z^{(*)})\times {\rm BR}(h^0\to b\overline b)\,.
\label{event_number}
\end{equation}
In the LHC searches, limits are often presented on the above product in which BR($Z^{(*)}\to \ell\overline \ell, \nu\overline \nu$) has been divided out.
We will calculate the signal cross section in eq.(\ref{event_number}) in the IH scenario in the 2HDM (Type I), and compare its magnitude with the corresponding cross section in the NH scenario ($m_{h^0}=125$ GeV), the latter being the current focus of the LHC searches in this channel.
In NH the product in eq.(\ref{event_number}) depends on three unknown parameters: $m_{A^0}$, $\tan\beta$ and $\cos(\beta-\alpha)$.
 In IH there is a fourth unknown parameter, $m_{h^0}$. The dependence of the three terms in eq.(\ref{event_number}) on the four unknown parameters is as follows (see also the discussion in section III):
 \begin{itemize}
\item[{(i)}]  The cross-section $\sigma(gg\to A^0)$ depends on $m_{A^0}$ and the couplings $A^0 t\overline t$ ($\propto \cot^2\beta$) and $A^0 b \overline b$ ($\propto\tan^2\beta$). Contributions from the 
couplings of $A^0$ to lighter fermions can be neglected due to their much smaller masses;
\item[{(ii)}]  BR$(A^0\to h^0Z^{(*)})$ is given by $\Gamma(A^0\to h^0Z^{(*)})/\Gamma^{total}_{A^0}$. 
The partial width $\Gamma(A^0\to h^0Z^{(*)})$ depends on $m_{A^0}$, the mass difference $m_{A^0}-m_{h^0}$ (in the phase space factor) and $\cos^2(\beta-\alpha)$ (in the square of the $A^0h^0Z$ coupling).
The total width $\Gamma^{total}_{A^0}$ is equal to  $\Gamma(A^0\to h^0Z^{(*)})+\Gamma^{rest}_{A^0}$, where $\Gamma^{rest}_{A^0}$ is the sum of the partial decay widths of 
all the other decays of $A^0$;
\item[{(iii)}]   BR$(h^0\to b\overline b$) given by $\Gamma(h^0\to b\overline b)/\Gamma^{total}_{h^0}$.
The partial width $\Gamma(h^0\to b\overline b)$ depends on $m_{h^0}$ and $\cos^2(\beta-\alpha)$ (e.g. via the coupling $\sin\alpha/\cos\beta$ in Type II and $\cos\alpha/\sin\beta$ in Type I). 
The total width $\Gamma^{total}_{h^0}$ is equal to  $\Gamma(h^0\to b\overline b)+\Gamma^{rest}_{h^0}$, where $\Gamma^{rest}_{h^0}$ is the sum of the partial decay widths of 
all the other decays of $h^0$. 
\end{itemize}
In what follows, numerical results for each of the three terms in eq.(\ref{event_number}) will be shown. Finally, we show the magnitude of the product of the three terms (i.e. the number of signal events) as a function
of $m_{A^0}$ in both IH (for various values of $m_{h^0}$) and NH, fixing the remaining parameters in the 2HDMs under consideration. All experimental and theoretical constraints in Section II are respected.
 In Fig.~\ref{fig:h0BR} to Fig.~\ref{fig:sigmaBR400} 
the parameter $m_{12}$ is taken to be $m_{12}^{2}=m_{h^0}^{2}(\frac{  \tan\beta }{1+\tan^2\beta})$, which ensures compliance with the experimental and theoretical constraints for the
chosen values and parameter ranges of the other 2HDM parameters. In Fig.~\ref{fig:sigmaBR170} we take $m_{12}^{2}=1000$ GeV$^{2}$ for the same reasons.
The BRs of $h^0$ and $A^0$ are calculated using 2HDMC \cite{Eriksson:2009ws}. 
We remark that we sampled only the portions of parameter space wherein the contribution of the channel $gg\to A^0\to h^0Z^{(*)}$ (in the narrow width approximation of $A^0$) is in close agreement with the yield of the full process $gg\to 
h^0Z^{(*)}$  (which also has contributions that do not involve $A^0$ i.e. $Z^{*} $ $s$-channel mediation and box diagrams at the amplitude level \cite{Accomando:2020vbo}). A study of the remainder of the parameter space using the latter process will be the subject of a future study.

In Fig.~\ref{fig:h0BR} the BRs of $h^0$ (i.e. the third term in the event number in eq.~(\ref{event_number})) in the 2HDM (Type I) are displayed 
as a function of $m_{h^0}$ in IH ($m_{H^0}=125$ GeV) with $\cos(\beta -\alpha )=1$, $\tan\beta=5.2$ and $m_{A^0}=m_{H^\pm}=140$ GeV.
The displayed range of values of $m_{h^0}$ is $40\,{\rm GeV} < m_{h^0} < 100$ GeV. In the 2HDM (Type I) the couplings $h^0f\overline f$ are scaled by
a factor of $\cos\alpha/\sin\beta$ relative to the couplings of the SM Higgs boson to the fermions, while the couplings $h^0WW$ and $h^0ZZ$ are scaled by $\sin(\beta-\alpha)$.
We take $\cos(\beta-\alpha)=1$ (which is an approximate requirement in IH due to the LHC measurements of the 125 GeV boson, interpreted as being $H^0$) and thus one has BR$(h^0\to WW)=0$ and BR$(h^0\to ZZ)=0$ at tree-level. Taking values of $\cos(\beta-\alpha)$ slightly less than 1 
(which is allowed from the measurements of $H^0$)
would give non-zero BR$(h^0\to WW)$ and BR$(h^0\to ZZ)$, but both channels would be very suppressed by the small value of $\sin^2(\beta-\alpha)$ and also by the phase space in the range of interest $40\,{\rm GeV} < m_{h^0} < 100$ GeV.
In Fig.~\ref{fig:h0BR} it can be seen that BR($h^0\to b\overline b)$ is around 90\%, and slightly decreases as $m_{h^0}$ increases towards $m_{h^0}=100$ GeV.
These values of BR$(h^0\to b\overline b)$ are larger than BR$(H^0\to b\overline b)\approx 58\%$ for the 125 GeV boson decaying to $b\overline b$.
The channel $h^0\to \tau^+\tau^-$ has the second-largest BR, being around 10\%. BR$(h^0\to gg)$ increases with $m_{h^0}$, with BR$(h^0\to \tau^+\tau^-)\approx {\rm BR}(h^0\to gg)$ for $m_{h^0}=100$ GeV.
The reason for this increase is due to the partial width $\Gamma(h^0\to gg)\propto m^3_{h^0}$ while $\Gamma(h^0\to b\overline b, \tau^+\tau^-)\propto m_{h^0}$. Other decay channels ($h^0\to c\overline c, \gamma\gamma, \gamma Z,{\rm etc}$)
have much smaller BRs and are not shown.

\begin{figure}
    \centering
    \includegraphics[width=17cm]{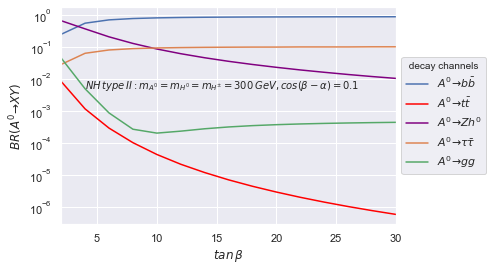}
    \caption{The BRs of $A^0$ in the 2HDM (Type II) as a function of $\tan\beta$ in the NH ($m_{h^0}=125$ GeV) with $\cos(\beta -\alpha )=0.1$ and $m_{A^0}=m_{H^0}=m_{H^\pm}=300$ GeV.}
    \label{fig:BRANHII}
\end{figure}

\begin{figure}
    \centering
    \includegraphics[width=17cm]{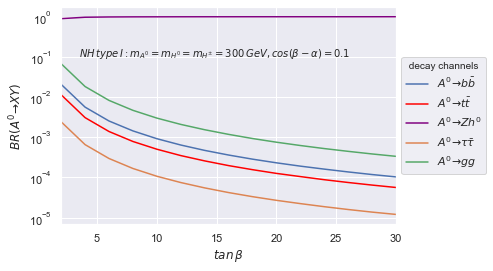}
    \caption{The BRs of $A^0$ in the 2HDM (Type I) as a function of $\tan\beta$ in the NH ($m_{h^0}=125$ GeV) with $\cos(\beta -\alpha )=0.1$ and $m_{A^0}=m_{H^0}=m_{H^\pm}=300$ GeV.}
    \label{fig:BRANHI}
\end{figure}

\begin{figure}
    \centering
    \includegraphics[width=17cm]{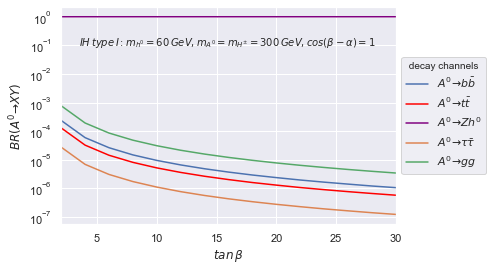}
    \caption{The BRs of $A^0$ in the 2HDM (Type I) as a function of $\tan\beta$ in the IH ($m_{H^0}=125$ GeV) with $\cos(\beta -\alpha )=1$ and $m_{A^0}=m_{H^\pm}=300$ GeV.}
    \label{fig:BRAIHI}
\end{figure}

In Fig.~\ref{fig:BRANHII} to Fig.~\ref{fig:BRAIHI} the BRs of $A^0$ (i.e. the second term in the event number in eq.~(\ref{event_number}))
as a function of $\tan\beta$ in three different scenarios are studied. In Fig.~\ref{fig:BRANHII} 
the BRs of $A^0$ are displayed in the 2HDM (Type II) as a function of $\tan\beta$ in the NH ($m_{h^0}=125$ GeV) with $\cos(\beta -\alpha )=0.1$ and $m_{A^0}=m_{H^0}=m_{H^\pm}=300$ GeV.
Five channels which can reach a BR of greater than 1\% are plotted, while channels that always have a smaller BR than $1\%$ are not plotted (although these would be present on the plot because the $y$-axis reaches BR$=10^{-6})$.
It can be seen from Fig.~\ref{fig:BRANHII} 
that $A^0\to h^0 Z$ of interest to this work has the largest BR (despite a suppression factor of $\cos^2(\beta-\alpha)=0.01$) until around $\tan\beta=3$, at which point $A^0\to b\overline b$ becomes the dominant decay due its partial width being proportional to $\tan^2\beta$ in the 2HDM (Type II).
The partial width of $A^0\to \tau^+\tau^-$ is also proportional to $\tan^2\beta$, and thus this decay becomes the second-most important channel for larger values of $\tan\beta$, reaching BR$(A^0\to \tau^+\tau^-)\approx 10\%$.
BR$(A^0\to h^0 Z)$ falls below $10\%$ for $\tan\beta>10$.
 BR($A^0\to gg$) is always less than a few percent and BR($A^0\to t\overline t$) (with one $t$ being virtual for the chosen value of $m_{A^0}=300$ GeV) is always less than 1\%.

Fig.~\ref{fig:BRANHI} is the same as Fig.~\ref{fig:BRANHII} (i.e. still NH) but for $A^0$ of the 2HDM (Type I). One can see that BR$(A^0\to h^0Z)$ is over $90\%$ for $\tan\beta\approx 3$ and is essentially $100\%$ for $\tan\beta>5$.
All other displayed channels have partial widths proportional to $\cot^2\beta$ and thus have increasingly small BRs (in contrast to Type II) as $\tan\beta$ increases.
Fig.~\ref{fig:BRAIHI} is the same as Fig.~\ref{fig:BRANHI} (i.e. for $A^0$ of the 2HDM (Type I)) but for IH. In Fig.~\ref{fig:BRAIHI}, three of the input parameters are changed, now being
$m_{H^0}=125$ GeV, $m_{h^0}=60$ GeV and $\cos(\beta -\alpha )=1$. The remaining two parameters are unchanged, being $m_{A^0}=m_{H^{\pm}}=300$ GeV. The larger value of 
$\cos(\beta -\alpha)$ and the smaller value of $m_{h^0}$ with respect to Fig.~\ref{fig:BRANHI}  means that BR$(A^0\to h^0Z)$ 
is even more dominant in IH than in NH, being essentially 100\% over the whole range of $\tan\beta$. The choice of $m_{A^0}=300$ GeV in Fig.~\ref{fig:BRANHI} and Fig.~\ref{fig:BRAIHI} ensures that the decay $A^0\to h^0 Z$ is a two-body decay, but
even for a virtual $Z^*$ (corresponding to lighter values of $m_{A^0}$) the magnitude of BR($A^0\to h^0 Z^*$) can be dominant. This will be apparent in later figures for the number of signal events
in eq.~(\ref{event_number}) which consider $m_{A^0}$ as low as 130 GeV.

\begin{figure}
    \centering
    \includegraphics [width=17cm]{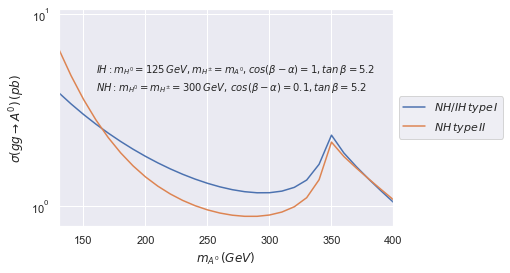}
    \caption{The cross section $\sigma(gg\to A^0)$ as a function of $m_{A^0}$ for NH with Type I, NH with Type II, and IH with Type I.
     The values of the input parameters are
    displayed on the figure, and  $m_{h^0}=55$ GeV, 75 GeV and 95 GeV in IH.}
    \label{fig:sigmaA}
\end{figure}

In Fig.~\ref{fig:sigmaA} the cross section $\sigma(gg\to A^0)$ (i.e. the first term in the event number in eq.~(\ref{event_number})) is displayed as a function of $m_{A^0}$ for NH with Type I, NH with Type II, and IH with Type I. 
The code Sushi \cite{Harlander:2012pb} is used to calculate $\sigma(gg\to A^0)$.
In NH the input parameters are $m_{H^0}=m_{H^\pm}=300$ GeV, $\cos(\beta-\alpha)=0.1$ and $\tan\beta=5.2$. 
In IH the input parameters are $m_{A^0}=m_{H^\pm}$, $\cos(\beta-\alpha)=1$, $\tan\beta=5.2$, and $m_{h^0}=55$ GeV, 75 GeV, 95 GeV. The cross section $\sigma(gg\to A^0)$ 
only depends on two 2HDM parameters, $m_{A^0}$ and $\tan\beta$ (as discussed in section IIIB) and in a given 2HDM its value is independent of NH or IH (because these two scenarios differ in $m_{h^0}, m_{H^0}$ and
$\cos(\beta-\alpha)$).
 Hence the lines for NH and IH in the 2HDM (Type I) coincide
and do not depend on the choice of $m_{h^0}$ in IH. The numerical difference in $\sigma(gg\to A^0)$ in the 2HDMs Type I and Type II arises from the fact that the coupling $A^0 b\overline b\propto \tan\beta$ in Type II 
and $A^0 b\overline b\propto \cot\beta$ in Type I, as shown in Table~\ref{2HDMAcoup}.  In Type I the top-quark loop contribution is essentially dominant. In contrast, in Type II the bottom-quark loop contribution is closer in magnitude to the
top-quark loop for the chosen value of $\tan\beta=5.2$ and interferes destructively, leading to a smaller cross section for 170 GeV$< m_{A^0} < 350$ GeV in Type II. In both models there is a local enhancement 
of $\sigma(gg\to A^0)$ at around $m_{A^0}=2m_t$, due to the $t$ quarks in the loop becoming on-shell. The magnitude of  $\sigma(gg\to A^0)$ is of the order of a few pb in the displayed range of 130 GeV$ < m_{A^0}< 400$ GeV. 

We are now ready to present the novel results of this work. In Fig.~\ref{fig:sigmaBR400} (upper panel)
the signal cross section $\sigma(gg\to A^0)\times {\rm BR}(A^0\to h^0Z^{(*)})\times {\rm BR}(h^0\to b\overline b)$ in eq.~(\ref{event_number}) is plotted as a function of $m_{A^0}$  
for NH with Type I, NH with Type II and for three choices of $m_{h^0}$  (55 GeV, 75 GeV, 95 GeV) in IH with Type I. The 2HDM input parameters are the same as in
 Fig.~\ref{fig:sigmaA}. In Fig.~\ref{fig:sigmaBR400} (lower panel), BR$(A^0\to h^0Z^{(*)})$ is plotted in IH only for the same range of $m_{A^0}$ and input parameters as in  Fig.~\ref{fig:sigmaBR400} (upper panel).
 It is essentially BR$(A^0\to h^0Z^{(*)})$ that determines the dependence of the signal cross section in Fig.~\ref{fig:sigmaBR400} (upper panel).
 Our results for the 2HDM Type I and Type II in NH in Fig.~\ref{fig:sigmaBR400} (upper panel)  agree with those presented in the LHC searches for $A^0\to h^0 Z$ (e.g. the CMS search in \cite{CMS:2019qcx},) with Type I having the larger
  signal cross section due to its larger BR$(A^0\to h^0Z^{(*)})$.
  Current searches at the LHC (for NH only) in this channel are sensitive to $m_{A^0}>225$ GeV. For  $m_{A^0}< 225$ GeV in NH the signal cross section starts to drop more sharply, the reason being that the $Z$ boson in the decay
 $A^0\to h^0 Z$ becomes off-shell for $m_{A^0}<216$ GeV. 
 
 We now compare the signal cross section for the 2HDM (Type I) in NH and IH. 
 It can be seen from Fig.~\ref{fig:sigmaBR400} (upper panel) that the signal cross section in NH Type I is similar in magnitude
 to that in IH Type I for  230 GeV$< m_{A^0}<330$ GeV. For these values of $m_{A^0}$ it can be seen from Fig.~\ref{fig:sigmaBR400} (lower panel) that BR$(A^0\to h^0Z^{(*)})$ is essentially 100\% in both IH and NH, 
 and $\sigma(gg\to A^0)$ is the same in both IH and NH for the 2HDM (Type I). The difference in the signal cross section solely arises from the fact that BR$(h^0\to b\overline b)\approx 85\%$ in IH while  
 BR$(h^0\to b\overline b)\approx 58\%$ in NH. For $m_{A^0}>330$ GeV one can see from Fig.~\ref{fig:sigmaBR400} (upper panel) that the signal cross section in IH becomes considerably larger than that in NH. This
 is because of the decreasing BR$(A^0\to h^0Z^{(*)})$ in NH (due to $\cos^2(\beta-\alpha)=0.01$ suppression in its partial width) as $A^0\to t\overline t$ gains in importance for $m_{A^0}>330$ GeV.    

Of most interest is the region $m_{A^0}<225$ GeV for which the current LHC searches (in NH only) have no sensitivity. For $m_{A^0}<225$ GeV
the signal cross section is much larger for IH, being around 1.2 pb for $m_{A^0}=150$ GeV and $m_{h^0}=95$ GeV, and increasing to 2.5 pb for 
 $m_{A^0}=150$ GeV and $m_{h^0}=55$ GeV. The reason for the much larger signal cross sections in IH is the fact that the $Z$ boson in the decay $(A^0\to h^0Z^{(*)})$ does not become off-shell until
 $m_{A^0}=146$ GeV, 166 GeV and 186 GeV for $m_{h^0}=55$ GeV, 75 GeV and 95 GeV respectively. This effect can be seen in Fig.~\ref{fig:sigmaBR400} (upper panel) in which the
 signal cross section starts to flatten as the $Z$ boson starts to become off-shell. We do not plot the signal cross section in IH for the other three 2HDMs with NFC (Type II, Lepton Specific and Flipped), which would
 have a smaller cross section than Type I. As mentioned earlier, the LHC searches set limits on all four 2HDMs in NH.
  
 In Fig.~\ref{fig:sigmaBR170} the signal cross section $\sigma(gg\to A^0)\times {\rm BR}(A^0\to h^0Z^{(*)})\times {\rm BR}(h^0\to b\overline b)$ as a function of $m_{A^0}$  
 is again displayed for NH with Type I, NH with Type II and for three choices of $m_{h^0}$ in IH with Type I. However, some input parameters are changed
 with respect to Fig.~\ref{fig:sigmaBR400} (upper panel). In Fig.~\ref{fig:sigmaBR170} we take 
 $\tan\beta=3$, 130 GeV$< m_{A^0}< 170$ GeV and the three values of $m_{h^0}$ in IH are 40 GeV, 70 GeV and 90 GeV. Moreover, the parameter
 $m^2_{12}$ is changed from its value in all previous figures ($=m_{h^0}^{2}(\frac{\tan\beta }{1+\tan^2\beta})$) to  $m^2_{12}=1000$ GeV$^2$ in order to comply with
 theoretical and experimental constraints. For the above choice of input parameters there are no valid points for $m_{A^0}>170$ GeV. The lower value of $\tan\beta$ gives rise to larger signal cross sections
 than in Fig.~\ref{fig:sigmaBR400} (upper panel), up to around 10 pb.

 In Table~\ref{BPinput} some benchmark points in the 2HDM (Type I) and IH are shown for $\tan\beta$ in the interval 2.9 to 5, with three of the points (BP1, BP2, BP3) being in the mass range
 80 GeV$<m_{A^0}+m_{h^0}<$110 GeV. In  Fig.~\ref{fig:sigmaBR400} and  Fig.~\ref{fig:sigmaBR170} the lowest value of $m_{A^0}+m_{h^0}$ was 170 GeV, but as discussed in Section III and in \cite{Moretti:2022fot},
 valid (experimentally unexcluded) points in the 2HDM (Type I) in IH can be found in the mass range 80 GeV$<m_{A^0}+m_{h^0}<$110 GeV. In Table \ref{BPoutput} the
 signal cross sections are presented, with numerical values reaching a few pb.
 
As discussed in section III, in  \cite{Moretti:2022fot} the mechanism $gg\rightarrow H^{0}\rightarrow A^{0}Z^{*}\rightarrow h^{0}Z^{*}Z^{*}\rightarrow b\bar{b}\mu ^{+}\mu ^{-}jj$
was proposed as a probe of the region 80 GeV$<m_{A^0}+m_{h^0}<$110 GeV. In Table~ \ref{moreBP} the signal cross section of the mechanism in \cite{Moretti:2022fot} 
is shown together with $\sigma(gg\rightarrow A^{0}\rightarrow h^{0}Z^{*}\rightarrow b\bar{b}\mu ^{+}\mu^{-})$, in which we now include the subsequent decay $Z^{*}\rightarrow \mu ^{+}\mu^{-}$ in order to
compare with the numerical values of the cross sections given in \cite{Moretti:2022fot}. It can be seen that 
 $\sigma(gg\rightarrow A^{0}\rightarrow h^{0}Z^{*}\rightarrow b\bar{b}\mu ^{+}\mu^{-})$ can be two orders of magnitude greater than that of 
 $\sigma(gg\rightarrow H^{0}\rightarrow A^{0}Z^{*}\rightarrow h^{0}Z^{*}Z^{*}\rightarrow b\bar{b}\mu ^{+}\mu ^{-}jj)$, and this is mainly due to the suppression factor
 of BR$(H^0\to A^0Z^*)\approx 0.2\%$. The experimental signatures are different, with $gg\rightarrow H^{0}\rightarrow A^{0}Z^{*}\rightarrow h^{0}Z^{*}Z^{*}\rightarrow b\bar{b}\mu ^{+}\mu ^{-}jj$
 having a smaller SM background due to the greater particle multiplicity of the signal. However, we expect $gg\rightarrow A^{0}\rightarrow h^{0}Z^{*}\rightarrow b\bar{b}\mu ^{+}\mu^{-}$
 to be a competitive probe of this region 80 GeV$<m_{A^0}+m_{h^0}<$110 GeV.

\newpage


    

\begin{figure}
\centering
  \centering
  \includegraphics[width=17cm]{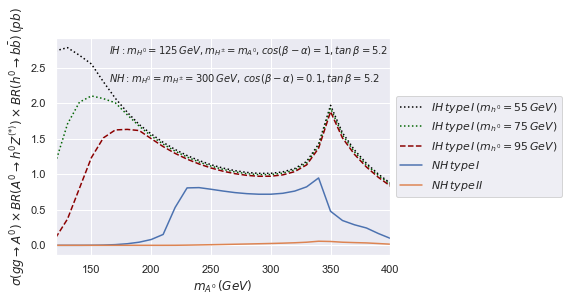}
  
  \label{fig:sub1}
  \centering
  \includegraphics[width=17cm]{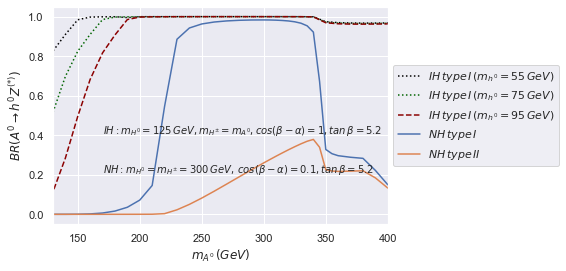}
  
  \label{fig:sigmaBR400}
\caption{Upper panel: the signal cross section $\sigma(gg\to A^0)\times {\rm BR}(A^0\to h^0Z^{(*)})\times {\rm BR}(h^0\to b\overline b)$ as a function of $m_{A^0}$  
for NH with Type I, NH with Type II and for three choices of $m_{h^0}$ in IH with Type I.  The values of the input parameters are
 displayed on the figure. \\
 Lower panel: Same as upper panel but for BR$(A^0\to h^0 Z^{(*)})$ alone. } 
  \label{fig:sigmaBR400}
\end{figure}

\begin{table}[h]
\begin{center}
\begin{tabular}{||c|c|c|c|c|c||} 
 \hline
 BP & $m_{A^{0}}$ & $m_{h^{0}}$ & $m_{H^{\pm }}$ & $\tan\beta$ & $\cos(\beta -\alpha )$ \\ [0.5ex] 
 \hline\hline
 1 & 80 & 12&80&4 & 1.0 \\ 
 \hline
2 & 93 & 15& 93&3.8 & 1.0 \\ 
 \hline
 3 & 75 & 10& 75 & 5 & 1.0  \\ 
 \hline
 4 & 155 & 80&155&2.9 & 1.0 \\ 
 \hline
 5 & 120 & 60&120&2.9& 1.0 \\ 
 \hline
 6 & 140 & 100&140&3 & 1.0   \\ 
 \hline
 7 & 100 & 90&100 & 3 & 1.0  \\ 
 \hline
\end{tabular}
\end{center}
\caption{Input parameters in 2HDM (Type I) and IH for 7 benchmark points. }
\label{BPinput}
\end{table}

\begin{table}[h]
\begin{center}
\begin{tabular}{||c|c|c|c|c||} 
 \hline
 BP &$\sigma (gg\rightarrow A^{0})_{\rm NNLO}$[pb] & BR$(A^{0}\to h^{0} Z^{(*)})$ & BR$(h^{0}\rightarrow b\bar{b})$ & $\sigma \times {\rm BR}(A^{0}\to h^{0} Z^{(*)})\times {\rm BR} (h^{0}\rightarrow b\bar{b})$[pb]\\ [0.5ex] 
 \hline\hline
 1 & 15.81&0.526&0.689& 5.72 \\ 
 \hline
 2 & 13.13&0.678 &0.804& 7.16 \\ 
 \hline
 3 &11.47&0.570&0.252& 1.64 \\ 
 \hline
 4  & 8.49 &0.592 &0.844& 4.24  \\ 
 \hline
 5  & 13.80 &0.336&0.861& 3.99 \\ 
 \hline
 6 & 9.61 & 0.070 & 0.823 & 0.55 \\ 
 \hline
 7 &18.36&0.00014&0.834&0.0021 \\ 
 \hline
\end{tabular}
\end{center}
\caption{Signal cross sections in 2HDM (Type I) and IH for the 7 benchmark points in Table \ref{BPinput}.}
\label{BPoutput}
\end{table}

\begin{table}[h]
\begin{center}
\begin{tabular}{||c| c|c||} 
 \hline
 BP & $\sigma(gg\rightarrow H^{0}\rightarrow A^{0}Z^{*}\rightarrow h^{0}Z^{*}Z^{*}\rightarrow b\bar{b} \mu ^{+}\mu ^{-}jj)\,$[pb] &$\sigma(gg\rightarrow A^{0}\rightarrow h^{0}Z^{*}\rightarrow b\bar{b} \mu ^{+}\mu^{-})\,$[pb]\\ [0.5ex] 
 \hline\hline
 8 (BP2 \cite{Moretti:2022fot})&$4.11\times 10^{-4} $ &0.105 \\ 
 \hline
9 (BP7 \cite{Moretti:2022fot})&$1.71\times 10^{-4} $ &0.141 \\ 
 \hline
 10 (BP24 \cite{Moretti:2022fot})& $3.54\times 10^{-4} $& $7.27\times 10^{-4}$\\ 
 \hline
 11 (BP10 \cite{Moretti:2022fot})&$3.31\times 10^{-4} $ &$5.48\times 10^{-2}$\\ 
 \hline
 12 (BP22 \cite{Moretti:2022fot})&$4.58\times 10^{-4} $ & $9.80\times 10^{-2}$ \\ 
 \hline
 13 (BP12 \cite{Moretti:2022fot})&$1.42\times 10^{-4} $& $9.90\times 10^{-2}$\\ 
 \hline
 14 (BP13 \cite{Moretti:2022fot})&$1.63\times 10^{-4}$& $9.02\times 10^{-2}$\\ 
 \hline
\end{tabular}
\end{center}
\caption{Comparison of signal cross sections for the mechanisms $\sigma(gg\rightarrow H^{0}\rightarrow b\bar{b}\mu ^{+}\mu ^{-}jj)$  in \cite{Moretti:2022fot} 
and $\sigma(gg\rightarrow A^{0}\rightarrow b\bar{b}\mu ^{+}\mu^{-} )\,$ in this work, as a probe of the region $m_{h^0}+m_{A^0}<110$ GeV, 
for some benchmark points in \cite{Moretti:2022fot}.}
\label{moreBP}
\end{table}

\section{Conclusions}  
\noindent
In this work we have studied the magnitude of the cross section for the production mechanism  $gg\to A^0 \to h^0 Z^{(*)}$ for a CP-odd scalar $A^0$ in the context of the 2HDM (Type I and II) in NH and 2HDM (Type I) in IH.
Current searches in this channel at the LHC are carried out assuming NH and take advantage of the measured mass $m_{h^0}=125$ GeV in order to optimise selection cuts and reduce the backgrounds to the signatures
$h^0\to b\overline b$ or $h^0\to \tau^+\tau^-$. In the absence of any signal, limits on the parameter space of $[\tan\beta, \cos(\beta-\alpha), m_{A^0}]$ in four types of 2HDM with NFC 
are derived for $m_{A^0} > 225$ GeV (i.e. for $A^0\to h^0 Z$ with an on-shell $Z$ boson).

Our novel results are for the scenario of IH in which $m_{H^0}=125$ GeV and $m_{h^0}$ is an unknown parameter that was varied in the range 10 GeV$ < m_{h^0} < 100$ GeV.
It was shown that the cross section for signal events $\sigma(gg\to A^0)\times {\rm BR}(A^0\to h^0Z^{(*)})\times {\rm BR}(h^0\to b\overline b)$ in the 2HDM (Type I) can be of the order of a few pb in IH
for the experimentally unexplored region of $m_{A^0} < 225$ GeV. Such cross sections are much larger than in NH, the reason being that BR$(A^0\to h^0 Z^{(*)})$ can stay large (even close to $100\%$)
for lower values of $m_{A^0}$ due to\\
 i) $m_{h^0}$ being smaller than 125 GeV, which keeps $Z$ on-shell to lower values of $m_{A^0}$, and ii) there being almost no suppression in the $A^0h^0Z$ coupling due to $\cos(\beta-\alpha)\approx 1$ in IH.

A signal for $A^0\to h^0 Z$ in IH would allow for simultaneous discovery of two Higgs bosons in the 2HDM. The current search strategy for $gg\to A^0 \to h^0 Z^{(*)}$ (which assumes NH) would need to
be slightly modified by removing the present cut of 100 GeV $ <m_{b\overline b} <140$ GeV on the invariant mass $m_{b\overline b}$ of the $b\overline b$ pair originating from the decay of $h^0$.
This cut could be replaced with smaller values of $m_{b\overline b}$ in order to capture most of the $b\overline b$ pairs from a light $h^0$
 in the range 10 GeV$ < m_{h^0} < 100$ GeV. We encourage a study (especially for $m_{A^0}<225 $ GeV) by the ATLAS/CMS collaborations of the detection prospects of the decay $A^0\to h^0 Z^{(*)}$ in the IH scenario.

\section*{Acknowledgements}
\noindent
SA acknowledges the use of the IRIDIS High Performance Computing Facility, and associated support services at the University of Southampton. SA acknowledges
support from a scholarship of the Imam Mohammad Ibn Saud Islamic University.
AA and SM are funded in part through the STFC CG ST/L000296/1.
SM is funded in part through the NExT Institute. We thank Souad Semlali for useful discussions.

\end{document}